\documentclass[a4paper,14pt]{article}
\usepackage{tabularx}
\usepackage{authblk}
\usepackage{amssymb}
\usepackage{amsthm}
\usepackage{amsmath}
\usepackage{setspace}
\usepackage{enumitem}
\usepackage{textcomp}

\usepackage[pdftex]{graphicx}
\usepackage[section]{placeins}
\usepackage[utf8]{inputenc}
\usepackage{wrapfig}
\usepackage{epstopdf}
\usepackage[toc,page]{appendix}
\usepackage{enumitem}

\newcommand{\beq}{\begin{equation}}
\newcommand{\eeq}{\end{equation}}

\begin{document}
\title{Conformal Standard Model and Inflation%
}
\author[${}^{1}$]{Jan Kwapisz}
\author[${}^{2}$]{Krzysztof A. Meissner}
\affil[${}^{1}$]{Institute of Mathematics,
Faculty of Mathematics, Informatics and Mechanics, University of Warsaw, Banacha 2, 02-097}
\affil[${}^2$]{Faculty of Physics, University of Warsaw,  Pasteura 5, 02-093 Warsaw, Poland}
\maketitle
\begin{abstract}
\noindent This article presents a possible inflation scenario as a consequence of non-minimal gravitational couplings in the Conformal Standard Model. The model consists, in comparison to the SM, of additional right-chiral neutrinos and complex scalars coupled to the right-chiral neutrinos but not to the SM particles. The inflation is driven by two non-minimally coupled fields one being the usual Higgs and the other one of the sterile scalars. It turns out that in this model the tensor to scalar ratio and spectral index can match the current data for a wide range of parameters. 
\end{abstract}
\section{Introduction}
It is usually assumed that at the very early stage of expansion of the Universe there was an inflation era i.e. an accelerated expansion followed by the decelerated FLRW epoch \cite{STAROBINSKY,Guthinflation}. This assumption is supposed to solve several outstanding cosmological problems but one should note that it has some problems of its own like the issue of its own initial conditions and the initial entropy problem. There were various mechanism proposed to provide the inflation era, see for example \cite{STAROBINSKY,Guthinflation,LINDE1982389, Freese,GARRIGA,Hawking:2000bb,1475-7516-2015-05-003,Bezrukov}. Large regions of parameters of these models were excluded by the recent BICEP and PLANCK measurements and the non-minimally coupled (to gravity) scalar field or modification of gravity by $R^2$ term are the main class of models left giving correct values for spectral index and tensor to scalar ratio.

The Higgs field is the only scalar field experimentally discovered so it is a natural candidate for an inflaton. In the Bezrukov, Shaposhnikov article \cite{Bezrukov} the consequences of Higgs field non-minimally coupled to gravity were considered and it turned out that it actually can give a successful inflation scenario. On the other hand the Standard Model struggles with some problems and an extended theory is required to explain both experimental phenomena not covered by SM itself and to resolve several high energy drawbacks of this theory.

Experimental data support such extensions which are in a sense minimal since there is no evidence of new (supersymmetric) particles from experiment so far. Such an extension is provided by the Conformal Standard Model (CSM) \cite{Meissner2007,Latosinski2015,Lewandowski2017} in which hierarchy problem is resolved by Softly Broken  Conformal Symmetry mechanism (SBCS) \cite{CSMtwo}, there is a candidate for dark matter and the observed matter-antimatter asymmetry can be explained. In this article we would like to point out that CSM supplemented by non-minimal coupling and eventually extended by one more scalar can also give a consistent description of inflation. Unlike the Higgs portal scenarios \cite{Wilczek, Lebedev:2011aq}, in CSM, the Higgs boson consists of two mass states. Moreover, the model predicts a tiny mass for minoron, which is a natural canditate for dark matter, produced during reheating. The Conformal Standard Model exists in two versions one with complex sextet $\phi_{ij}$ \cite{Latosinski2015} and one with additional complex scalar \cite{Lewandowski2017}, but as we will see in the next paragraph they both give the same scenario for inflation.

\section{Inflation scenario within Conformal Standard Model}
\label{CSMinflation}
In this paragraph we propose and discuss scalar fields from CSM non-minimally coupled to gravity leading to inflation in Bezrukov-Shaposhnikov mechanism \cite{Bezrukov}.  The potential is given by the formula
(we use the following notation: $h := H_0(x)$ and $s := r(x)$, compare with \cite{Latosinski2015,Lewandowski2017}.):
\begin{align}
\label{VCSMI}
V_J(h,s) = \frac{1}{4}\lambda_1(h^2-v_H^2)^2 + \frac{1}{4}\lambda_p(s^2-v_{\phi}^2)^2\nonumber\\
 + \frac{1}{2}\lambda_3\left(h^2-v_H^2\right)\left(s^2-v_{\phi}^2\right),
\end{align}
The conditions for the stability of this potential were discussed in \cite{Latosinski2015,Lewandowski2017}:
\beq
\label{first}
\begin{array}{lcr}
\lambda_1, \lambda_2, \lambda_4 >0, & & \lambda_3 > -\sqrt{\lambda_1(\lambda_2 + \lambda_4/3)}.
\end{array}
\eeq
Then the inflation comes from non-minimal coupling  to gravity i.e. the lagrangian (in the Jordan frame):
\beq
\mathcal{L} = \frac{1}{2}\partial_{\mu}h\partial^{\mu}h + \frac{1}{2}\partial_{\mu}s \partial^{\mu}s - \frac{\left(M_P^2 +\xi_1 h^2 +  \xi_2s^2\right)}{2} R- V_J(h, s),
\eeq
with $\xi_i>0$. In case of \cite{Latosinski2015}, because we take large couplings (see below), the $\phi_{ij}$ reduces to its trace part. We will proceed in the scheme of \cite{Lebedev:2011aq,Two-Higgs-doublet}. 
Since the calculations in the Jordan and Einstein frames are equivalent \cite{EinsteinJordan} we make the following convenient conformal transformation:
\beq
\label{CSMcoupling}
\begin{array}{lcr}
\tilde{g}_{\mu\nu}= \Omega^2 g_{\mu\nu}, & & \Omega^2 = 1 + \frac{\xi_1 h^2 + \xi_2s^2}{ M_P^2}.
\end{array}
\eeq
We set $M_P=1$ to simplify the equations but it will be restored later on for slow-roll parameters analysis. Then we obtain:
\beq
\label{LagrangianE}
\mathcal{L}_{\textrm{E}} = -\frac{R}{2} +\frac{3}{4}\left[\partial_{\mu}\log(\Omega^2)\right]^2 +\frac{1}{2\Omega^2}\left[(\partial_{\mu}h)^2 + (\partial_{\mu}s)^2\right] - \frac{1}{\Omega^4}V(H_0,r).
\eeq
Since in the inflation scenario we consider large fields limit (\cite{Multiconformal})  we take:
\beq
\xi_1h^2 + \xi_2s^2 \gg M_P^2 \gg v_i^2,
\eeq
We redefine the fields as: 
\begin{align}
\chi =& \sqrt{\frac{3}{2}}\log(\xi_1 h^2 + \xi_2s^2), \\
\tau =& \frac{h}{s},
\end{align}
and then the kinetic part of the lagrangian reads:
\begin{align}
\mathcal{L}_{\textrm{kin}} &= \frac{1}{2}\left(1 + \frac{1}{6}\frac{\tau^2 +1}{\xi_1\tau^2 + \xi_2}\right)(\partial_{\mu}\chi)^2 + \frac{1}{\sqrt{6}}\frac{(\xi_2 - \xi_1)\tau}{(\xi_1\tau^2 +\xi_2)^2}(\partial_{\mu}\chi)(\partial^{\mu}\tau) \\
&+ \frac{1}{2}\frac{\xi_1^2\tau^2 + \xi_2^2}{(\xi_1\tau^2 + \xi_2)^3}(\partial_{\mu}\tau)^2.
\end{align} 
We are interested in large fields and coupling regime: $\xi' = \xi_1 + \xi_2 \gg 1$. 
This is the same case as for single Higgs inflation, where $\xi \approx 49000\sqrt{\lambda}$. Then the mixing term $(\partial_{\mu}\chi)(\partial^{\mu}\tau)$ and the second term in front of $(\partial_{\mu}\chi)^2$ are suppressed by term $1/(\xi')$ so the kinetic part is:
\beq
\mathcal{L}_{\textrm{kin}} \simeq \frac{1}{2}(\partial_{\mu}\chi)^2 + \frac{1}{2}\frac{\xi_1^2\tau^2 +\xi_2^2}{(\xi_1\tau^2 + \xi_2)^3} (\partial_{\mu}\tau)^2,
\eeq  
and the potential in new variables reads:
\beq
\label{VECSM}
V_E(\tau, \chi)= U(\tau)W(\chi) = \frac{\lambda_1\tau^4 + \lambda_p+ 2\lambda_3\tau^2}{4(\xi_1\tau^2 + \xi_2)^2}\left(1 + e^{-2\chi/\sqrt{6}}\right)^{-2}.
\eeq
The calculated  minima of $U(\tau)$ are shown in the table below, with $a = \lambda_1 \xi_2 - \lambda_3 \xi_1$, $b = \lambda_p\xi_1 - \lambda_3\xi_2$. One can show that the ratio of fields drops eventually to stable minimum before the inflation ends and Shaposhnikov-type evolution of $\chi$ follows. We will discuss both issues below after we analyse the minima of $U(\tau)$ and parameters ensuring successful inflation.
\FloatBarrier
\begin{table}[h!]
\label{MinimaU}
\caption{Minimal values of the radial part of inflation potential}
\begin{tabular}{c|c|c}
$\tau_0$ values & stable minimum condition & $U_0$ \\
\hline
$\tau_0 =0$ & $a>0$ and $b<0$ & $\frac{\lambda_1}{4\xi_1^2}$, \\
$\tau_0 = + \infty$ & $a<0$ and $b>0$ & $\frac{\lambda_p}{4\xi_2^2}$, \\
$\tau_0 = \pm \sqrt{\frac{b}{a}}$ & $a>0$ and $b>0$ & $\frac{\lambda_1 \lambda_p - \lambda_3^2}{4(\lambda_1\xi_2^2 + \lambda_p \xi_1^2 - 2\lambda_3 \xi_1\xi_2)}$,\\
$\tau=0$ or $\tau_0= +\infty$ & $a<0$ and $b<0$ & $\frac{\lambda_1}{4\xi_1^2}$ or $\frac{\lambda_p}{4\xi_2^2}$.
\end{tabular}
\end{table}
\FloatBarrier
\noindent We have two types of scenarios. Either we have a single Inflaton case: Higgs or single ``shadow'' Higgs inflation, when $\tau_0$ is equal to zero or infinity. Otherwise the multi-inflaton scenario occurs where ratio of fields goes to the value: $\tau_0 = \sqrt{\frac{b}{a}}$. For multi-inflaton scenario the following conditions have to be satisfied: 
\begin{align}
a = \lambda_1 \xi_2 - \lambda_3 \xi_1>0 \label{positiveI},\\
b = \lambda_p\xi_1 - \lambda_3\xi_2 > 0 \label{positiveII},\\
\lambda_1\lambda_p - \lambda_3^2 >0  \label{positiveIII},
\end{align}
\noindent
where the third one is required to prevent metastability of electroweak vacuum and assure positivity of vacuum energy during the inflation stage. For $\lambda_3<0$ both: $a>0$ and $b>0$ are obviously satisfied. On the other hand the last condition is automatically satisfied for $\lambda_3>0$ since it comes from the first two.  Moreover, the choice $\lambda_3<0$ is more convenient to match the theoretical Higgs mass with its observed (125 GeV) value. So for this choice of parameters the CSM model is consistent with the inflation scenario. The measurement of $\lambda_3$ would be crucial to determine which of the final values of $\tau_0$ could be obtained in inflation. Assuming that the shadow Higgs was found in the LHC and the measured value of $\lambda_3$ was less than zero then the single Higgs scenario presented in \cite{Bezrukov} would be falsified even for $\xi_1 =0$ or $\xi_2 =0$. Moreover, for $\lambda_3>0$ single or mixed Higgs scenario can be realised in CSM with non-minimal couplings.  \\
In the the case $\xi_i=0$ for $i \in \{1,2\}$ the field associated with $i$'th coupling decouples \cite{Two-Higgs-doublet}, which results in $\tau = 0, +\infty$. When this is not the case, one can still show \cite{Two-Higgs-doublet,Lebedev:2011aq} that the $\tau$ field will drop to the one of the minima showed in the Tab.~[\ref{MinimaU}]. The easiest way to obtain this result is to use slow roll approximation for both fields and then solve the differential equations for evolution of $\tau$ field.\\
So we are left with classical Bezrukov-Shaposhnikov evolution:
\beq
V(\chi) = \frac{\lambda_{eff}}{4\xi^2}W(\chi),
\eeq
with $\xi = \xi_1\tau_0^2 + \xi_2$ and $\lambda_{eff} = \lambda_1 + \lambda_p\tau_0^4 + 2\lambda_3\tau_0^2$. Here we recall the general scheme \cite{Bezrukov}: for large $\chi$ potential is flat and the inflation occurs and  as the field rolls to smaller values the $\epsilon \simeq 1$ marks the end of inflation. The slow roll parameters are calculated below:
\begin{align}
\epsilon &= \frac{M_P^2}{2}\left(\frac{dW/d\chi}{W}\right)^2 \simeq \frac{4M_P^2}{3}\frac{e^{-4\chi/\sqrt{6}}}{\left(1+e^{-2\chi/\sqrt{6}}\right)^2},\\
\eta &= M_P^2 \frac{d^2W/d^2 \chi}{W} \simeq - \frac{4M_P^2}{3}e^{-2\chi/\sqrt{6}}\frac{1 - 2e^{-2\chi/\sqrt{6}}}{\left(1 + e^{-2\chi/\sqrt{6}}  \right)^2}.
\end{align}
The number of e-folds is given by:
\beq
N = \int_e^i \frac{V(\chi)}{V'(\chi)}d\chi = \frac{3}{4}\left[ e^{2\chi_i/\sqrt{6}} - e^{2\chi_e/\sqrt{6}} + \frac{2}{\sqrt{6}}\left(\chi_i - \chi_e\right)\right],
\eeq
and the slow roll conditions are violated for $e^{2\chi_e/\sqrt{6}} \simeq 0.155$. Then the initial value of the field, for $N=60$, is given by: $e^{2\chi_e/\sqrt{6}} \simeq 80$. Hence the initial values of fields, after the decoupling stage, are given by:
\begin{align}
s_i & \simeq M_P \sqrt{\frac{4N}{3\xi}}, \\
h_i & \simeq M_P \sqrt{\frac{4N}{3\xi_1}}\sqrt{\left[1 - \frac{\xi_2}{\xi}\right]}.
\end{align}
Inserting it into COBE normalisation \cite{Lyth:1998xn}: $W/\epsilon = (0.027M_P)^4$ and with $N_{COBE} \simeq 62$, we obtain (in analogy to Bezrukov-Shaposhnikov):
\beq
\xi \simeq \sqrt{\frac{\lambda_{eff}}{3}}\frac{N_{COBE}}{0.027^2} \simeq 49000 \sqrt{\lambda_{eff}},
\eeq
for $\lambda_{eff} \sim 1$ the coupling to gravity is: $\xi \simeq 49000 = \xi_1 \tau_0^2+ \xi_2$. Since $\tau^2 \simeq v_H^2/v_{\phi}^2 = \mathcal{O}(1)$, then roughly $\xi' \simeq \xi$. The spectral index is \cite{Bezrukov,Two-Higgs-doublet} (since it depends on the shape of the potential rather than its amplitude):
\beq
n_s \simeq 1 - \frac{2}{N} \simeq 0.97,
\eeq
with tensor to scalar ratio: $r \simeq 12/N^2 \simeq 0.0033$, what agrees with WMAP3 and Planck data.
\section{Conclusions and remarks}
\label{Conclusions}
The Conformal Standard Model with two scalar fields non-minimally coupled to the Ricci scalar is analysed in the scheme along the lines of \cite{Shaposhnikov:2006xi}. For both versions of the model non-minimal couplings provide the same inflation scenario in accordance with the observed values of the spectral index and the tensor to scalar ratio. Therefore the Conformal Standard Model with resonant leptogenesis mechanism \cite{Lewandowski2017,PILAFTSIS2004303} and the described inflation scenario can provide an explanation of the cosmological phenomena we observe.  In most models extensions of the Standard Model and inflation scenarios are treated separately. In contradistinction to previous proposals, our model incorporates inflation into CSM, which is a well established model. 
\\\\
We thank H. Nicolai for discussions. J.K. and K.A.M. thanks the Albert Einstein
Institute in Potsdam for hospitality and support during this work. K.A.M. was partially supported by the Polish National Science Center grant DEC-2017/25/B/ST2/00165.

\bibliographystyle{siam}
\end{document}